\documentclass[draftcls, onecolumn, 12pt]{IEEEtran}
\usepackage{threeparttable,placeins,makecell,stfloats,cite}
\usepackage{amsmath,amssymb}
\usepackage[dvips]{graphicx}
\usepackage{amsfonts}
\usepackage{latexsym}
\usepackage{graphics} % for pdf, bitmapped graphics files
\usepackage{epsfig} % for postscript graphics files

\usepackage{standalone}
\usepackage{tikz}
\usetikzlibrary{positioning}
\usetikzlibrary{decorations.markings}

\newtheorem{lemma}{\textit{Lemma}}

\newtheorem{example}{\textit{Example}}

\usepackage[format=hang]{subfig}
\ifCLASSINFOpdf
 \else
 \fi

\begin{document}

%\markboth{IEEE Signal Processing Letters, Vol. XX, No. Y, Month 2014} {Liu, Guan, and Hu: Optimal Spectrally Constrained Sequences \ldots}

%\title{Revisit to The Lower Bounds on The Aperiodic Correlation of Sequence Set Over The Complex Roots-of-Unity}
\title{CPM Training Waveforms With Autocorrelation Sidelobes Close To Zero}
\author{Zilong~Liu,~%\IEEEmembership{Student Member,~IEEE},
        Yong~Liang~Guan,~%~%,~%\IEEEmembership{Member,~IEEE}
        and~Chee-Cheon~Chui
\thanks{Zilong Liu was with the School
of Electrical and Electronic Engineering, Nanyang Technological
University, Singapore. He is now with Institute of Communication Systems, Home of 5G Innovation Centre, University of Surrey, UK (E-mail: {\tt zilong.liu@surrey.ac.uk}). Yong Liang Guan is with the School
of Electrical and Electronic Engineering, Nanyang Technological University, Singapore (E-mail: {\tt eylguan@ntu.edu.sg}). C.-C. Chui is with DSO National Laboratories, Singapore 609081 (e-mail: ccheeche@dso.org.sg).}}
 \maketitle

 \begin{abstract}
Continuous phase modulation (CPM) plays an important role in wireless communications due to its constant envelope signal property and tight spectrum confinement capability. Although CPM has been studied for many years, CPM training waveforms having autocorrelations with zero sidelobes have not been reported before, to the best of our knowledge. Existing works on CPM system design mostly assume that the channel fading coefficients are either \textit{perfectly} known at the receiver or estimated using random CPM training waveforms. In this work, we propose a novel class of CPM training waveforms displaying autocorrelation sidelobes close to zero. The key idea of our construction is to apply differential encoding to Golay complementary pair having perfect aperiodic autocorrelation sum properties.
\end{abstract}


\section{Introduction}

Continuous phase modulation (CPM) is an attractive nonlinear modulation scheme whose signals exhibit properties of constant envelope and tight spectrum confinement \cite{CPM-book}. The first property will allow the transmitter to enjoy high power transmission efficiency as CPM signals have peak-to-average power ratio (PAPR) of $1$ (theoretically). Hence, traditional transmission techniques (e.g., \cite{Jiang05,Jiang08}) to deal with high PAPR problem in orthogonal frequency-division multiplexing (OFDM) may be avoided when high-rate transmission is not a must. The second property implies less amount of out-of-band power leakage compared to 2- and 4-ary PSK modulations \cite{CPM-book}, therefore, causing less interference to other applications (operated over adjacent spectral bands) and leading to relatively higher spectral efficiency.

Nowadays, CPM has been used in many areas such as global system for mobile communications (GSM) \cite{GSMmod1997}, military and satellite communications \cite{EZB2012}, millimeter communications \cite{Heath2007}, and machine-type communications in 5G \cite{CPM-SCFDMA,MTC2016}. %For CPM transmission over flat-fading channels, maximum likelihood sequence detection (MLSD) of CPM signals can be efficiently implemented via Viterbi algorithm. A low-complexity receiver was proposed in \cite{Kalen1989} based on Laurent decomposition, which represents a binary CPM signal as superposition of a series of PAM pulses (called Laurent pulses) modulated by certain pseudo-symbols \cite{Laurent1986}.
In dispersive channels, frequency-domain equalization (FDE) is needed to suppress the effect of intersymbol interference (ISI), followed by CPM demodulation as in flat-fading channels. FDE based CPM receiver design can be found in \cite{Tan2005,Pancaldi2006,Thillo2009}, in which the channel fading coefficients are assumed to be \textit{perfectly} known at the receiver. Recently, \textit{random} sequence based channel estimation and equalization has been investigated in \cite{Chayot2017}.

%To achieve performance gain, CPM is usually serially concatenated with different channel codes (e.g., convolutional codes, BCH codes, and LDPC codes) \cite{BBB1999,NS2001,Amat_2009,Xiao_LCPM_2005,Benaddi_2014}. At the receiver, iterative (turbo) decoding is conducted by exchanging soft reliability information between the CPM (inner) decoder and the channel (outer) decoder \cite{Narayanan_1999}, where CPM can be demodulated by a maximum a posteriori (MAP) decoder implemented using BCJR algorithm \cite{BCJR1974}.

Despite of a long history of CPM research, less has been understood on the training waveform design of CPM. According to \cite{GSM-Modern-1999}, the 8 training binary sequences defined in GSM standard \cite{GSM1997} have been found by computer search over all possible $2^{16}$ binary sequences. In 2013, Hosseini and Perrins studied the training sequence design of burst-mode CPM over additive white Gaussian noise (AWGN) channels \cite{HP-TCOM-1,HP-TCOM-2}. However, the CPM training sequences proposed in \cite{HP-TCOM-1} may not be applicable in frequency-selective channels (as will be shown in Section IV). Motivated by this, we target at a systematic construction of CPM training waveforms for frequency-selective channels. Our main idea is to apply differential encoding to Golay complementary pair (GCP) whose aperiodic autocorrelation sums diminish to zero for all the non-zero time-shifts \cite{Golay61}. Taking advantage of Laurent decomposition \cite{Laurent1986}, we show that the resultant CPM training waveform (with modulation index of $h=1/2$) displays autocorrelation sidelobes close to zero.

\section{Preliminaries}

\subsection{Introduction to CPM}
Let $j=\sqrt{-1}$. An equivalent lowpass $M$-ary CPM waveform $s(t)$ is expressed as $s(t)=\exp\left [ j \phi(t;\mathbf{I})\right ]$,
%\begin{equation}\label{defi_CPM_sig}
%s(t)=\exp\left [ j \phi(t;\mathbf{I})\right ],
%\end{equation}
where $\phi(t;\mathbf{I})=2\pi h \sum_{k=0}^{n}I_k q(t-kT)$ ($nT\leq t \leq (n+1)T$)
%\begin{equation}
%\phi(t;\mathbf{I})=2\pi h \sum_{k=0}^{n}I_k q(t-kT),~~nT\leq t \leq (n+1)T,
%\end{equation}
is the time-varying phase depending on the information sequence $\mathbf{I}=\{I_k\}_{k=0}^{n}$ with $I_k$ being the $k$-th CPM symbol drawn from the set of $\left \{ \pm 1, \pm3, \cdots, \pm(M-1)\right \}$, $h$ is the modulation index, and $T$ is the symbol duration. The phase-shaping waveform $q(t)$ is defined as the integral of the frequency-shaping pulse $g(t)$ of duration $LT$, i.e., $q(t)=\int_{0}^{t}g(\tau) d\tau$,
%\begin{equation}
%q(t)=\int_{0}^{t}g(\tau) d\tau,
%\end{equation}
with $q(t)=0$ for $t\leq 0$ and $q(t)=1/2$ for $t\geq LT$. $s(t)$ is called full-response if $L=1$ and partial-response when $L>1$. Note that $\phi(t;\mathbf{I})$ can be written as
\begin{equation}
\phi(t;\mathbf{I})=2\pi h \sum\limits_{k=0}^{n-L}I_k q(t-kT)+2\pi h \sum\limits_{k=n-L+1}^{n}I_k q(t-kT).
\end{equation}
Let $\theta_n=\pi h \sum\limits_{k=0}^{n-L}I_k$ and $\sigma_n=[I_{n-1},I_{n-2},\cdots,I_{n-L+1}]$. It is easy to see that the phase $\phi(t;\mathbf{I})$ depends on the modulator state $\chi_n=[\theta_n,\sigma_n]$, where $\theta_n$ and $\sigma_n$ are called the \textit{phase state} and the \textit{correlative state}, respectively.

\noindent \textit{Laurent's Decomposition}: Laurent showed that a binary partial-response CPM signal can be represented as a superposition of a number of pulse-amplitude-modulated (PAM) pulses \cite{Laurent1986}. To introduce this, we first define
\begin{equation}
s_0(t)=
\begin{cases}
\frac{\sin 2\pi h q(t)}{\sin \pi h},~&0\leq t\leq LT,\\
\frac{\sin [\pi h - 2\pi h q(t-LT)]}{\sin \pi h},~&LT\leq t\leq 2LT,\\
0,~&\text{otherwise}.
\end{cases}
\end{equation}
Also, denote by $a_{p,m}\in\{0,1\}$ the coefficients of the binary representation of integer $p$ in the set of $\{0,1,\cdots,2^{L-1}-1\}$, i.e., $p=\sum\limits_{m=1}^{L-1}2^{m-1}a_{p,m}.$
%\begin{equation}
%p=\sum\limits_{m=1}^{L-1}2^{m-1}a_{p,m}.
%\end{equation}
Then, the CPM signal can be written as
\begin{equation}
s(t)=\sum\limits_{n}\sum\limits_{p=0}^{2^{L-1}-1}\exp\left [j\pi h A_{p,n}\right]\cdot c_p(t-nT),
\end{equation}
where $c_p(t)=s_0(t)\prod_{i=1}^{L-1}s_0[t+(i+La_{p,i})T]$,
%\begin{equation}
%c_p(t)=s_0(t)\prod_{i=1}^{L-1}s_0[t+(i+La_{p,i})T],
%\end{equation}
for $0\leq t \leq T\times \min\limits_{i=1,2,\cdots,L-1}[L(2-a_{p,i})-i]$, and $A_{p,n}=\sum\limits_{m=0}^{n}I_m-\sum\limits_{m=1}^{L-1}I_{n-m}a_{p,m}.$
%\begin{equation}
%A_{p,n}=\sum\limits_{m=0}^{n}I_m-\sum\limits_{m=1}^{L-1}I_{n-m}a_{p,m}.
%\end{equation}
In general, $c_0(t)$ is the most important PAM pulse as it carries more than $99\%$ of the total signal energy \cite{Laurent1986}. Therefore,
the CPM signal can be approximated as
\begin{equation}\label{appr_CPM_sig}
s(t)\approx \alpha(t)\triangleq\sum\limits_{n}\gamma_n\cdot c_0(t-nT),
\end{equation}
with
\begin{equation}\label{equ_gamma_n}
\gamma_n \triangleq \exp[j\pi h A_{0,n}]=\exp\left[j\pi h \cdot \sum\limits_{m=0}^{n}I_m\right ].
\end{equation}
For ease of presentation, $\{\gamma_n\}$ are called CPM pseudo-symbols.
\subsection{Introduction to Golay Complementary Pair (GCP)}
Denote by $\rho_{\textbf{C}}(k)$ the aperiodic auto-correlation function (AACF) of $\textbf{C}=[C_0,C_1,\cdots,C_{N-1}]$ which is defined as
\begin{equation}
\rho_{\textbf{C}}(k)=
\begin{cases}
\sum\limits_{n=0}^{N-1-k}C_n C^*_{n+k},~~&\text{if}~0\leq k\leq N-1;\\
\rho^*_{C}(-k),~~&\text{if}~1-N\leq k<0;\\
0,~~&\text{otherwise}.
\end{cases}
\end{equation}
Let $(\textbf{C},\textbf{D})$ be a pair of sequences with identical length of $N$. $(\textbf{C},\textbf{D})$ is called a GCP \cite{Golay61} if $\rho_{\textbf{C}}(k)+\rho_{\textbf{D}}(k)=0$ for any $k\neq0$. Note that compared to conventional one-dimensional sequences, the two constituent sequences in a GCP work in a cooperative way to ensure that their out-of-phase aperiodic autocorrelations sum to zero.

Let $\phi_{\textbf{C}}(k)=\sum_{k=0}^{N-1}C_kC^*_{n+k~\text{mod}~N}$ be the periodic autocorrelation function (PACF) of $\textbf{C}$ at time-shift $k$. Clearly, $\phi_{\textbf{C}}(k)+\phi_{\textbf{D}}(k)=0$ for any $k\neq0~(\text{mod}~N)$ if $(\textbf{C},\textbf{D})$ is a GCP.

Denote by $\mathbb{Z}_q$ the set of integers modulo $q$. For $\underline{x}=[x_1,x_2,\cdots,x_{\nu}]\in \mathbb{Z}_2^{\nu}$, a generalized Boolean function (GBF) $f(\underline{x})$ (or $f(x_1,x_2,\cdots,x_{\nu})$)
is defined as a mapping $f: \{0,1\}^{\nu} \rightarrow \mathbb{Z}_q$. Let
$(i_1,i_2,\cdots,i_{\nu})$ be the binary representation of the
integer $i=\sum_{k=1}^{\nu} i_k 2^{k-1}$, with $i_{\nu}$ denoting the most significant bit. Given $f(\underline{x})$ (or $f(x_1,x_2,\cdots,x_{\nu})$), define $f_i\triangleq f(i_1,i_2,\cdots,i_{\nu})$, and
\begin{displaymath}
\textit{\textbf{f}}\triangleq\Bigl [f(0,0,\cdots,0), f(1,0,\cdots,0), \cdots, f(1,1,\cdots,1) \Bigl ].
\end{displaymath}
%\vspace{0.1in}
We present the example below to illustrate GBFs defined above. One can find it useful in understanding the GCP construction in \textit{Lemma \ref{PSK_GDJ constr_4GCP}} (which is formed by summation of a series of quadratic and linear terms of GBFs).
\begin{example}
Let $\nu=3$ and $q=2$. The associated sequences of $1,x_1,x_3,x_1x_3$ are
\begin{displaymath}
\begin{array}{ccl}
\mathbf{1}   &=&(1,1,1,1,1,1,1,1),\\
\mathbf{x}_1 &=&(0,1,0,1,0,1,0,1),\\
\mathbf{x}_3 &=&(0,0,0,0,1,1,1,1),\\
\mathbf{x}_1\mathbf{x}_3+\textbf{1}&=&(1,1,1,1,1,0,1,0),
\end{array}
\end{displaymath}
respectively.
\end{example}

%\vspace{0.1in}
\begin{lemma} \label{PSK_GDJ constr_4GCP}(Davis-Jedwab Construction of GCP \cite{DAVIS-JEWAB99})
{Let}
\begin{equation}\label{f_4GDJ_GCP}
f(\underline{x})\triangleq \frac{q}{2} \sum \limits_{k=1}^{\nu-1}x_{\pi(k)} x_{\pi(k+1)} + \sum
\limits_{k=1}^{\nu} c_k x_k+c,
\end{equation}
where $\pi$ is a permutation of the set $\{1,2,\cdots,\nu\}$, and $c_k,c\in\mathbb{Z}_{q}$ ($q$ even integer).
Then, for any $c'\in \mathbb{Z}_{q}$, $\textit{\textbf{f}}~~\text{and}~~\textit{\textbf{f}}+\frac{q}{2}\textit{\textbf{x}}_{\pi(1)}+c'\cdot \textbf{1}$
%\begin{equation}
%\textit{\textbf{f}}~~\text{and}~~\textit{\textbf{f}}+\frac{q}{2}\textit{\textbf{x}}_{\pi(1)}+c'\cdot \textbf{1}
%\end{equation}
form a GCP over $\mathbb{Z}_{q}$ of length $2^\nu$. %$\xi=\exp(2\pi\sqrt{-1}/{2^h} )$.
\end{lemma}

%\vspace{0.1in}
%\begin{remark}
%A binary GCP is obtained by setting $q=2$ in \textit{Lemma \ref{PSK_GDJ constr_4GCP}}.
%\end{remark}
%\vspace{0.1in}
%\vspace{0.1in}

\section{Proposed CPM Training Waveform Design}
In this section, we will propose a training waveform design for CPM signal $s(t)$ [or the approximation $\alpha(t)$] with periodic autocorrelation sidelobes close to zero.
Throughout the proposed design, we consider binary CPM with $h=1/2$. Therefore, $\exp[j\pi h]=j$. To get started, we first consider a sequence of $\{\gamma_n\}_{n=0}^{2N-1}$ with $2N$ non-zero elements satisfying $N> L+1$ and
\begin{equation}\label{periodic_trans}
\gamma_n=\gamma_{n+N},~\text{for}~0\leq n \leq N-1.
\end{equation}
The PACF of the CPM approximation $\alpha(t)$ [cf. (\ref{appr_CPM_sig})] over $[NT,2NT]$ is defined as
\begin{equation}
\phi_{\alpha}(\tau)=\int_{t=NT}^{2NT}\alpha(t)\alpha^*(t+\tau)dt,
\end{equation}
where
\begin{equation}
\begin{split}
\alpha(t+\tau)=
\begin{cases}
\alpha(t+\tau-NT),& ~\text{if}~t+\tau>2NT;\\
\alpha(t+\tau+NT),&~\text{if}~t+\tau<NT.
\end{cases}
\end{split}
\end{equation}
%Similarly, the PACF of $\{\gamma_n\}_{n=0}^{N-1}$ with respect to an integer time-shift $k$ is defined as
%\begin{equation}
%\phi_{\gamma}(k)=\sum\limits_{n=0}^{N-1}\gamma_n\gamma^*_{n+k~\text{mod}~N}.
%\end{equation}
%Denote by $\rho_{\gamma}(k)$ the aperiodic auto-correlation function (AACF) of $\{\gamma_n\}_{n=0}^{N-1}$ which is defined as
%\begin{equation}
%\rho_{\gamma}(k)=
%\begin{cases}
%\sum\limits_{n=0}^{N-1-k}\gamma_n\gamma^*_{n+k},~~&\text{if}~0\leq k\leq N-1;\\
%\rho^*_{\gamma}(-k),~~&\text{if}~1-N\leq k<0;\\
%0,~~&\text{otherwise}.
%\end{cases}
%\end{equation}
%It is easy to see that
%\begin{equation}
%\phi_{\gamma}(k)=\rho_{\gamma}(k)+\rho^*_{\gamma}(N-k).
%\end{equation}
In addition, the AACF of $c_0(t)$, the most significant pulse in Laurent's decomposition, is defined as
\begin{equation}\label{AACF_c0}
\rho_{c_0}(\tau)=\int_{t=-\infty}^{+\infty}c_0(t)c_0(t+\tau)dt,
\end{equation}
where $c_0(t)=0$ if $t<0$ or $t>(L+1)T$. Hence, $\rho_{c_0}(\tau)=0$ if $|\tau|>(L+1)T$.
By [\ref{CDMA-handBook}, (6.45b)], we have
\begin{equation}\label{PACF_alpha}
\phi_{\alpha}(\tau)=\sum\limits_{k=-\infty}^{+\infty}\phi_{\gamma}(k)\rho_{c_0}(\tau-kT).
\end{equation}
(\ref{PACF_alpha}) implies that the PACF of $\alpha(t)$ [i.e., $\rho_{\alpha}(\tau)$] is zero for any $|\tau|\geq (L+1)T$ if
\begin{equation}\label{equ_perfect_gamma}
\phi_{\gamma}(k)=0,~~\forall~k\neq 0.
\end{equation}
%Nevertheless, it may not be straightforward to find sequence $\{\gamma_n\}$ which satisfies (\ref{equ_gamma_n}) and (\ref{equ_perfect_gamma}) simultaneously. Inspired by this, we introduce Golay complementary pairs (GCPs), each consisting of a pair of sequences whose AACFs (and PACFs) sum to zero for any non-zero time-shifts.

%\subsection{Proposed Design}

\begin{figure}%[htbp]
  \centering
  %\captionsetup{justification=centering}
  %\includegraphics[width=4.5in]{CPM-Training-Seq.pdf}\\
  \includegraphics[width=3.5in]{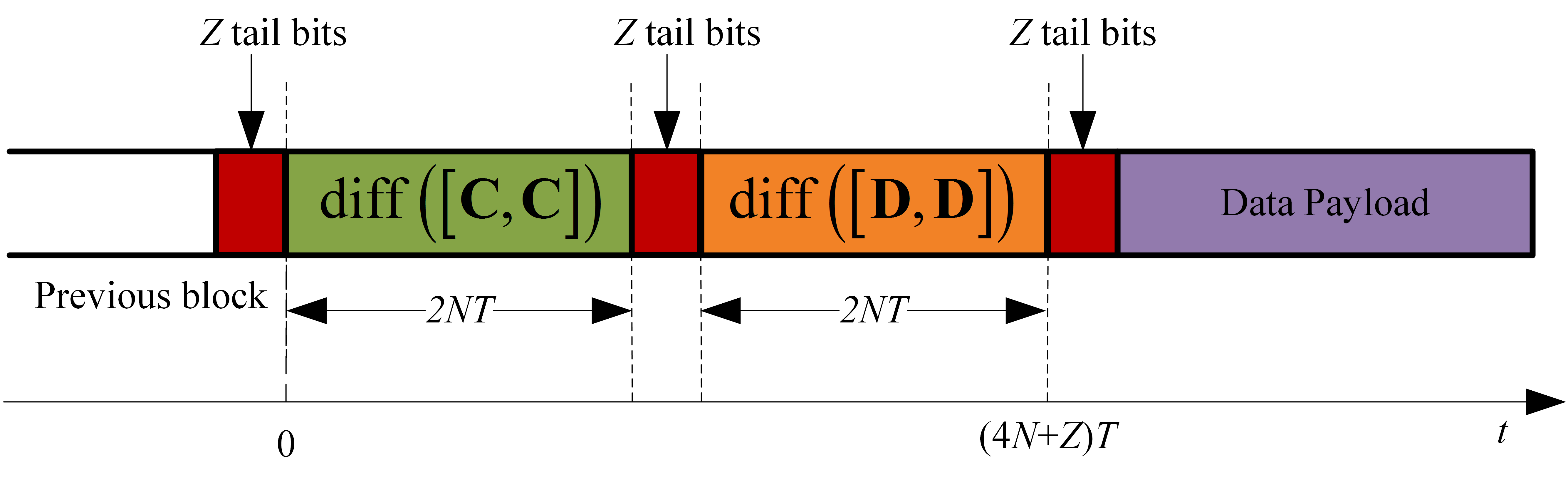}\\
  \caption{CPM block structure consisting of training sequence and data payload.}
  \label{CPM-train-seq}
\end{figure}
Nevertheless, it is hard to find sequence $\{\gamma_n\}$ satisfying (\ref{equ_gamma_n}) and (\ref{equ_perfect_gamma}) simultaneously. Because of this, we consider sequence pair $\textbf{C}=[C_0,C_1,\cdots,C_{N-1}]$ and $\textbf{D}=[D_0,D_1,\cdots,D_{N-1}]$ over $\{0,1\}^N$. Applying the ``differential-encoding", denoted by $\text{diff}(\cdot)$, to $\tilde{\textbf{C}}\triangleq[\textbf{C},\textbf{C}]$, we obtain $I_{\textbf{C}} \triangleq\text{diff}([\textbf{C},\textbf{C}])=[I_{C,0},I_{C,1},\cdots,I_{C,2N-1}]$
%\begin{displaymath}
%\begin{split}
%I_C & \triangleq\text{diff}([C,C])=[I_{C,0},I_{C,1},\cdots,I_{C,2N-1}],%\\
%%I_D & \triangleq\text{diff}([D,D])=[I_{D,0},I_{D,1},\cdots,I_{D,2N-1}].
%\end{split}
%\end{displaymath}
with
\begin{equation}
\begin{split}
I_{C,m} & =(2\tilde{C}_m-1)\cdot (2\tilde{C}_{m-1}-1)\\
         & =4\tilde{C}_m\tilde{C}_{m-1}-2\tilde{C}_m-2\tilde{C}_{m-1}+1\in\{-1,1\},
\end{split}
\end{equation}
%and
%\begin{equation}
%\begin{split}
%I_{D,m} & =(2\tilde{D}_m-1)\cdot (2\tilde{D}_{m-1}-1)\\
%         & =4\tilde{D}_m\tilde{D}_{m-1}-2\tilde{D}_m-2\tilde{D}_{m-1}+1\in\{-1,1\},
%\end{split}
%\end{equation}
where $0\leq m \leq 2N-1$ and $\tilde{C}_{-1}=1$. It is clear that
\begin{equation}
\exp[\pi h I_{C,m}] =j \cdot (-1)^{\tilde{C}_m} \cdot (-1)^{\tilde{C}_{m-1}}.
\end{equation}
Similarly, applying ``differential-encoding" to $\tilde{\textbf{D}}\triangleq[\textbf{D},\textbf{D}]$, we obtain $I_{\textbf{D}} \triangleq\text{diff}([\textbf{D},\textbf{D}])=[I_{D,0},I_{D,1},\cdots,I_{D,2N-1}]$
%\begin{displaymath}
%\begin{split}
%%I_C & \triangleq\text{diff}([C,C])=[I_{C,0},I_{C,1},\cdots,I_{C,2N-1}],%\\
%I_D & \triangleq\text{diff}([D,D])=[I_{D,0},I_{D,1},\cdots,I_{D,2N-1}],
%\end{split}
%\end{displaymath}
assuming $\tilde{D}_{-1}=1$. As we will see later, differential-encoding helps to mitigate the correlations among CPM symbols which facilitates the design of CPM training waveforms with autocorrelation sidelobes close to zero. Then, send $I_{\textbf{C}}$ and $I_{\textbf{C}}$ for CPM modulation following the transmission structure shown in Fig. \ref{CPM-train-seq}. The ``$Z$ tail bits", placed before $I_{\textbf{C}}$ (or $I_{\textbf{D}}$), are used to return the CPM modulator phase state to zero. Systematic method for the generation of tail bits can be found in \cite{Tan2005} using Rimoldi decomposition \cite{Rimoldi1988}. %For ease of presentation, suppose the length of tail bits is $Q$.

Applying $I_{\textbf{C}}$ as a sequence of $2N$ CPM symbols, the CPM pseudo-symbols can be expressed as follows.
\begin{enumerate}
\item If $n\in \{0,1,2,\cdots,N-1\}$, we have
\begin{equation}
\begin{split}
\gamma_{C,n} & =\exp\left [j \pi h \cdot \sum\limits_{m=0}^{n}I_{C,m} \right ] \\
 & =j^{n+1} \cdot (-1)^{C_n}\cdot (-1)^{C_{-1}}= j^{n+3} \cdot (-1)^{C_n}.
% & = j^{n+3} \cdot (-1)^{C_n}.
\end{split}
\end{equation}

\item If $n\in \{N,N+1,N+2,\cdots,2N-1\}$, we have
\begin{equation}
\begin{split}
\gamma_{C,n} & =\exp\left [j \pi h \cdot \sum\limits_{m=0}^{n}I_{C,m} \right ] \\%= j^{n-N+3} \cdot (-1)^{C_{n-N}}\cdot j^N.
& = j^{n-N+3} \cdot (-1)^{C_{n-N}}\cdot j^N.
\end{split}
\end{equation}
\end{enumerate}

When $N\equiv 0~\text{mod}~4$, one can see that
\begin{equation}\label{periodic_trans2}
[\gamma_{C,0},\gamma_{C,1},\cdots,\gamma_{C,N-1}]=[\gamma_{C,N},\gamma_{C,N+1},\cdots,\gamma_{C,2N-1}].
\end{equation}
In this case, (\ref{periodic_trans}) holds. This implies that $I_{\textbf{C}}$ is a periodic transmission of two identical length-$N$ sequences, which in turn allows the calculation of PACF at the local receiver. The approximated CPM waveform $\alpha_{\textbf{C}}(t)$ in (\ref{appr_CPM_sig}) can be written as
\begin{equation}\label{equ_alpha_a}
\alpha_{\textbf{C}}(t)= \sum\limits_{n=0}^{2N-1}\underbrace{j^{n+3}\cdot (-1)^{C_n}}_{\gamma_{C,n}}\cdot c_0(t-nT),~\text{for}~0\leq t \leq 2NT.
\end{equation}
%Now, we consider sequence $\mathbf{b}$ over $\{0,1\}^N$ which, together with $\mathbf{a}$, form a binary GCP.
Denote by $\alpha_{\textbf{D}}(t)$ the approximated CPM waveform (after ``differential encoding") corresponding to $I_{\textbf{D}}$. Similar to (\ref{periodic_trans2}), we have
\begin{equation}\label{periodic_trans3}
[\gamma_{D,0},\gamma_{D,1},\cdots,\gamma_{D,N-1}]=[\gamma_{D,N},\gamma_{D,N+1},\cdots,\gamma_{D,2N-1}],
\end{equation}
for $N\equiv 0~\text{mod}~4$. Also, similar to (\ref{equ_alpha_a}), we obtain
\begin{equation}\label{equ_beta_b}
\begin{split}
\alpha_{\textbf{D}}(t) & = \sum\limits_{n=0}^{2N-1}\underbrace{j^{n+3}\cdot (-1)^{D_n}}_{\gamma_{D,n}}\cdot c_0(t-nT-2NT-ZT),%\\
%            & ~~~~~~~~~~~~~~~~~~~~~~~~~~~~~~~~~~~~~~\text{for}~2NT+MT\leq t \leq 4NT+MT.
\end{split}
\end{equation}
for $2NT+ZT\leq t \leq 4NT+ZT$.

Next, we will show $\gamma_{\textbf{C}}\triangleq\left\{\gamma_{C,n}\right \}_{n=0}^{N-1}$ and $\gamma_{\textbf{D}}\triangleq\left \{\gamma_{D,n}\right \}_{n=0}^{N-1}$ form a quaternary GCP provided that $(\textbf{C},\textbf{D})$ is a binary GCP generated by the Davis-Jedwab construction (see \textit{Lemma \ref{PSK_GDJ constr_4GCP}}). In the context of \textit{Lemma \ref{PSK_GDJ constr_4GCP}}, let
\begin{equation}
f=\sum \limits_{k=1}^{\nu-1}x_{\pi(k)} x_{\pi(k+1)} + \sum
\limits_{k=1}^{\nu} c_k x_k+c~~(\text{mod}~2),
\end{equation}
and $f+x_{\pi(1)}+c'~(\text{mod}~2)$ be the GBFs of $\textbf{C}$ and $\textbf{D}$, respectively, where $N=2^{\nu}$ ($\nu\geq2$ as $N$ should be divisible by 4). Lifting these two GBFs from $\mathbb{Z}_2$ to $\mathbb{Z}_4$ and noting that $n=\sum_{k=1}^{\nu} x_k 2^{k-1}$, the corresponding GBFs of $\gamma_{C,n}$ and $\gamma_{D,n}$ can be expressed as
\begin{equation}
\begin{split}
f_{\textbf{C}} & =2\sum \limits_{k=1}^{\nu-1}x_{\pi(k)} x_{\pi(k+1)} + 2\sum
\limits_{k=1}^{\nu} c_k x_k\\
& ~~~~+2c+\sum_{k=1}^{\nu} x_k 2^{k-1}+3~~(\text{mod}~4)\\
   & =2\sum \limits_{k=1}^{\nu-1}x_{\pi(k)} x_{\pi(k+1)} + 2\sum
\limits_{k=1}^{\nu} c_k x_k\\
   & ~~~~+2c+2x_2+x_1+3~~(\text{mod}~4)\\
   & =2\sum \limits_{k=1}^{\nu-1}x_{\pi(k)} x_{\pi(k+1)} + \sum
\limits_{k=3}^{\nu} (2c_k) x_k\\
   & ~~~~+(2+2c_2)x_2+(1+2c_1)x_1+2c+3~~(\text{mod}~4),
\end{split}
\end{equation}
and $f_{\textbf{D}}=f_{\textbf{C}}+2x_{\pi(1)}+2c'~(\text{mod}~4)$, respectively. It is clear that $f_{\textbf{C}},f_{\textbf{D}}$ satisfy the GBF forms in \textit{Lemma \ref{PSK_GDJ constr_4GCP}}. Thus, $\gamma_{\textbf{C}}$ and $\gamma_{\textbf{D}}$  are a quaternary GCP.
Applying (\ref{PACF_alpha}) to (\ref{equ_alpha_a}) and (\ref{equ_beta_b}), we assert that
\begin{equation}
\begin{split}
   & \phi_{\alpha_{\textbf{C}}}(\tau)+\phi_{\alpha_{\textbf{D}}}(\tau)\\
 = &\sum\limits_{k=-\infty}^{+\infty}\left [\phi_{\gamma_{\textbf{C}}}(k)+\phi_{\gamma_{\textbf{D}}}(k)\right ]\rho_{c_0}(\tau-kT)\\
 = &\begin{cases}
2NT\rho_{c_0}(\tau),& ~\text{if}~0\leq|\tau|\leq(L+1)T;\\
0, & ~\text{if}~(L+1)T<|\tau|\leq (N-L-1)T.
\end{cases}
\end{split}
\end{equation}
Consider the CPM training waveform over $\{s(t):0\leq t\leq (4N+Z)T\}$ and let
\begin{equation}
\begin{split}
S^{(1)}_{\textbf{C}} & =\{s(t):0\leq t\leq NT\}, \\
S^{(2)}_{\textbf{C}} & =\{s(t):NT\leq t\leq 2NT\},\\
S^{(1)}_{\textbf{D}} & =\{s(t):(2N+Z)T\leq t\leq (3N+Z)T\},\\
S^{(2)}_{\textbf{D}} & =\{s(t):(3N+Z)T\leq t\leq (4N+Z)T\}.
\end{split}
\end{equation}
At the receiver, $S^{(2)}_{\textbf{C}}$ and $S^{(2)}_{\textbf{D}}$ are taken as two local reference waveforms for correlation with $S^{(1)}_{\textbf{C}}$ and $S^{(1)}_{D}$, respectively. This is because $S^{(1)}_{\textbf{C}}$ (and $S^{(1)}_{\textbf{D}}$) will be spread into the time window of $NT\leq t\leq 2NT$  [and $(3N+Z)T\leq t\leq (4N+Z)T$] owing to the multipath propagation. Finally, we have the following assertion.
\begin{equation}
\begin{split}
         & \phi_{S^{(1)}_{\textbf{C}},S^{(2)}_{\textbf{C}}}(\tau)+\phi_{S^{(1)}_{\textbf{D}},S^{(2)}_{\textbf{D}}}(\tau)\\
 \approx &~\phi_{\alpha_{\textbf{C}}}(\tau)+\phi_{\alpha_{\textbf{D}}}(\tau)\\
 \approx &~0, ~\text{if}~(L+1)T<|\tau|\leq (N-L-1)T.
\end{split}
\end{equation}

\section{Simulation Results}

\begin{figure}%[htbp]
  \centering
  \includegraphics[trim=3cm 9.5cm 3cm 9.5cm, clip=true, width=3.5in]{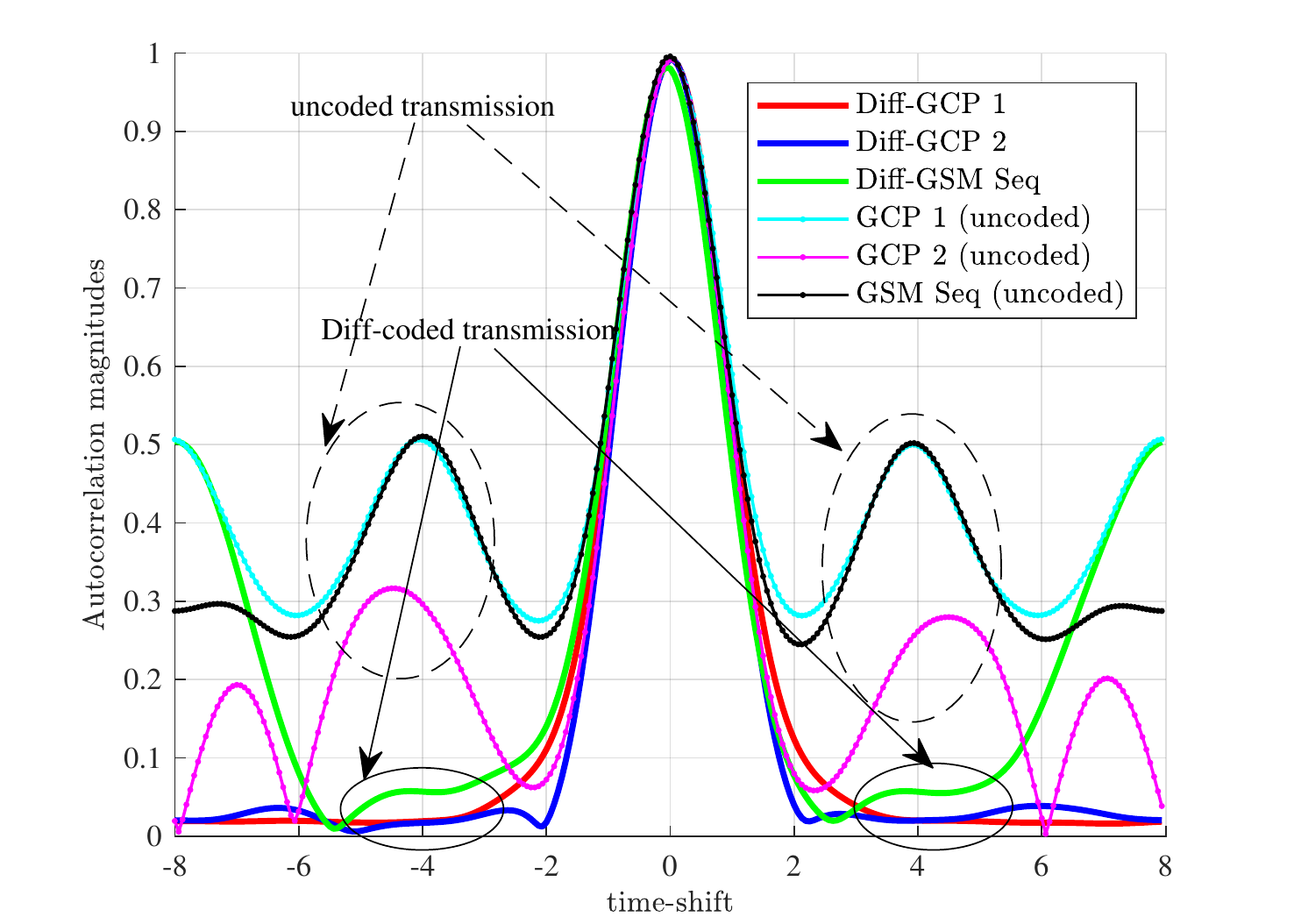}\\
  \caption{Autocorrelation comparison of different CPM waveforms}
  \label{CPM-wav-comp}
\end{figure}

In the context of \textit{Lemma \ref{PSK_GDJ constr_4GCP}}, let $q=2,\nu=4$ (i.e., $N=2^\nu=16$) and $c=0,c'=1$. Consider two GCPs, with $\pi=[1,2,3,4]$ and $[c_1,c_2,c_3,c_4]=[1,0,1,1]$ for GCP 1, and $\pi=[2,3,4,1]$ and $[c_1,c_2,c_3,c_4]=[1,1,0,1]$ for GCP 2. The resultant GCPs are given below.
\begin{displaymath}
\text{GCP 1}=\left[ \begin{matrix}%{cc}
+-++-+++-+---+++\\
---+++-++++-++-+
          \end{matrix} \right],
\end{displaymath}
\begin{displaymath}
\text{GCP 2}=\left[ \begin{matrix}
+--+-+-+++--++++\\
-+-++--+------++
          \end{matrix} \right].
\end{displaymath}
Applying differential encoding to GCPs 1 and 2, we obtain differentially-encoded pairs, denoted by ``Diff-GCP 1" and ``Diff-GCP 2", respectively. Each pair will be sent as $I_{\textbf{C}}$ and $I_{\textbf{D}}$ (see Section III) for CPM modulation following the transmission structure in Fig. \ref{CPM-train-seq}. For comparison, differential encoding is also applied to GSM sequence $[+-+++----+---+--]$ \cite{GSM1997}. The resultant sequence is referred to as ``Diff-GSM" and will be sent as $I_{\textbf{C}}$ only for CPM training.
For simulation, we consider the GMSK frequency pulse $g(t)$ below.
\begin{displaymath}
g(t)=\frac{1}{2T}\left [ Q\left (\frac{\frac{t}{T}+\frac{1}{2}}{\sigma}\right ) - Q\left (\frac{\frac{t}{T}-\frac{1}{2}}{\sigma}\right )\right ],
\end{displaymath}
where $Q(t)=\int_{t}^{+\infty}\frac{1}{\sqrt{2\pi}}\exp\left ( -\frac{x^2}{2}\right )dx$ and $\sigma^2=\frac{\text{In}2}{4\pi^2(BT)^2}$.
%\begin{displaymath}
%Q(t)=\int_{t}^{+\infty}\frac{1}{\sqrt{2\pi}}\exp\left ( -\frac{x^2}{2}\right )dx,~~\sigma^2=\frac{\text{In}2}{4\pi^2(BT)^2}.
%\end{displaymath}
Considering GMSK frequency pulse with $BT=0.3$ and truncated frequency pulse length of $L=3$, we obtain CPM training waveform $\{s(t):0\leq t\leq (4N+Z)T\}$, where $Z=3$ in this example. The (normalized) autocorrelation magnitudes of CPM waveforms, i.e., $\left |\phi_{S^{(1)}_{\textbf{C}},S^{(2)}_{\textbf{C}}}(\tau)+\phi_{S^{(1)}_{\textbf{D}},S^{(2)}_{\textbf{D}}}(\tau)\right |$, are shown in Fig. \ref{CPM-wav-comp}. One can see that the proposed training waveforms 1 and 2 (corresponding to Diff-GCPs 1 and 2, respectively) exhibit autocorrelation sidelobes close to zero for time-shifts larger than $(L+1)T$. In contrast, the training waveform from the Diff-GSM sequence exhibits considerably large autocorrelation sidelobes when time-shift is larger than $6T$. To show the effectiveness of differential-encoding, we also depict the autocorrelation magnitudes of uncoded GCPs 1 and 2, and uncoded GSM sequence. It is shown that CPM waveforms after differential-encoding exhibit lower autocorrelation sidelobes compared to the uncoded ones.

%\begin{table*}
%\small
%\centering
%\tabcolsep=0.11cm
%\caption{Differentially-encoded CPM sequences}
%%\resizebox{\columnwidth}{!}{
%
%\begin{tabular}{|c||c|}
%\hline
%    & Training Sequence   \\ \hline \hline
%Diff-GCP 1  & $\left( \begin{matrix}
%+--+--++---++-+++--+--++---++-++\\
%-++-++--+++--+---++-++--+++--+--
%          \end{matrix} \right) $  \\  \hline \hline
%Diff-GCP 2 & $ \left ( \begin{matrix}
%+-+-----++-+-++++-+-----++-+-+++\\
%----+-+--+++++-+----+-+--+++++-+
%    \end{matrix} \right) $   \\  \hline \hline
%Diff-GSM sequence & $\left ( \begin{matrix}
%+--++-+++--++--+---++-+++--++--+\\
%    \end{matrix} \right) $   \\  \hline
%\end{tabular}
%%}
%\end{table*}

\begin{figure}%[htbp]
  \centering
  \includegraphics[trim=3cm 9.5cm 3cm 9.5cm, clip=true, width=3.5in]{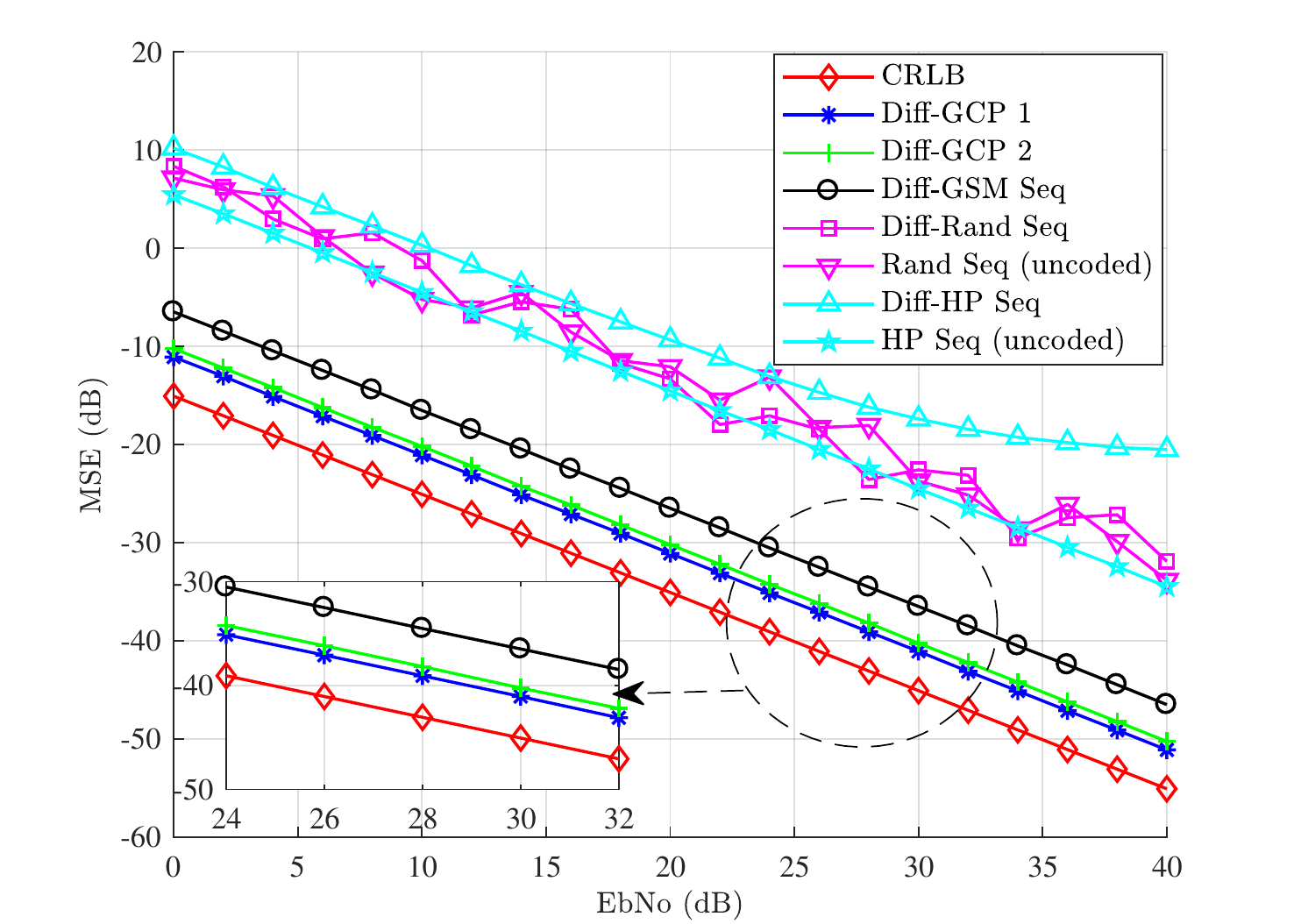}\\
  \caption{Comparison of channel estimation MSEs using different CPM training waveforms}
  \label{CPM-ChanEst-Comp}
\end{figure}

Next, we apply a set of CPM training waveforms, which are normalized to have \textit{identical energy} in the transmission, for estimation of a 16-path channel (separated by integer symbols duration) having uniform power delay profile. Specifically, we consider $h[t]=\sum_{n=0}^{15}h_i\delta[t-nT]$, where $h_i$'s are complex-valued Gaussian random variables with zero mean and $\mathbb{E}(|h_i|^2)=1/16$. These CPM training waveforms are generated based on Diff-GCPs 1 and 2, Diff-GSM, uncoded random sequences, ``Diff-Rand" sequences (i.e., differentially-encoded random ``on-the-fly" sequences), uncoded Hosseini-Perrins (HP) sequence $[----++++++++----]$ \cite{HP-TCOM-1}, and ``Diff-HP" sequence. Using least squares (LS) estimator, comparison of channel estimation mean-squared-errors (MSEs) of different CPM training waveforms is shown in Fig. \ref{CPM-ChanEst-Comp}. The CPM training waveforms from uncoded random sequences, Diff-Rand sequences, uncoded HP sequence, and Diff-HP sequence result in relatively higher MSEs due to their autocorrelation sidelobes with larger variations and rank-deficient (sometimes) LS estimator. The GSM training waveform leads to MSE performance 8.5dB away from the Cram\'{e}r-Rao lower bound (CRLB)\footnote{This CRLB is derived for perfect CPM training waveform with zero autocorrelation sidelobes for all the non-zero time-shifts \cite{Milewski1983}.}. The MSEs of using the proposed training waveforms (based on differentially encoded GCPs) exhibit MSEs much closer to CRLB, as close as 4dB from the CRLB for the case of ``Diff-GCP 1" (compared to 5dB distance for ``Diff-GCP 2"). This is understandable as the training waveform from Diff-GCP 1 displays the best autocorrelation performance with uniformly low sidelobes, as shown in Figure \ref{CPM-wav-comp}. It should be noted that the estimates of $h_1$ and $h_{15}$ are more likely to suffer from larger MSEs due to the roll-off autocorrelation sidelobes at $\pm T$ (see Figure \ref{CPM-wav-comp}).

\begin{figure}%[htbp]
  \centering
  \includegraphics[trim=3cm 9.5cm 3cm 9.5cm, clip=true, width=3.5in]{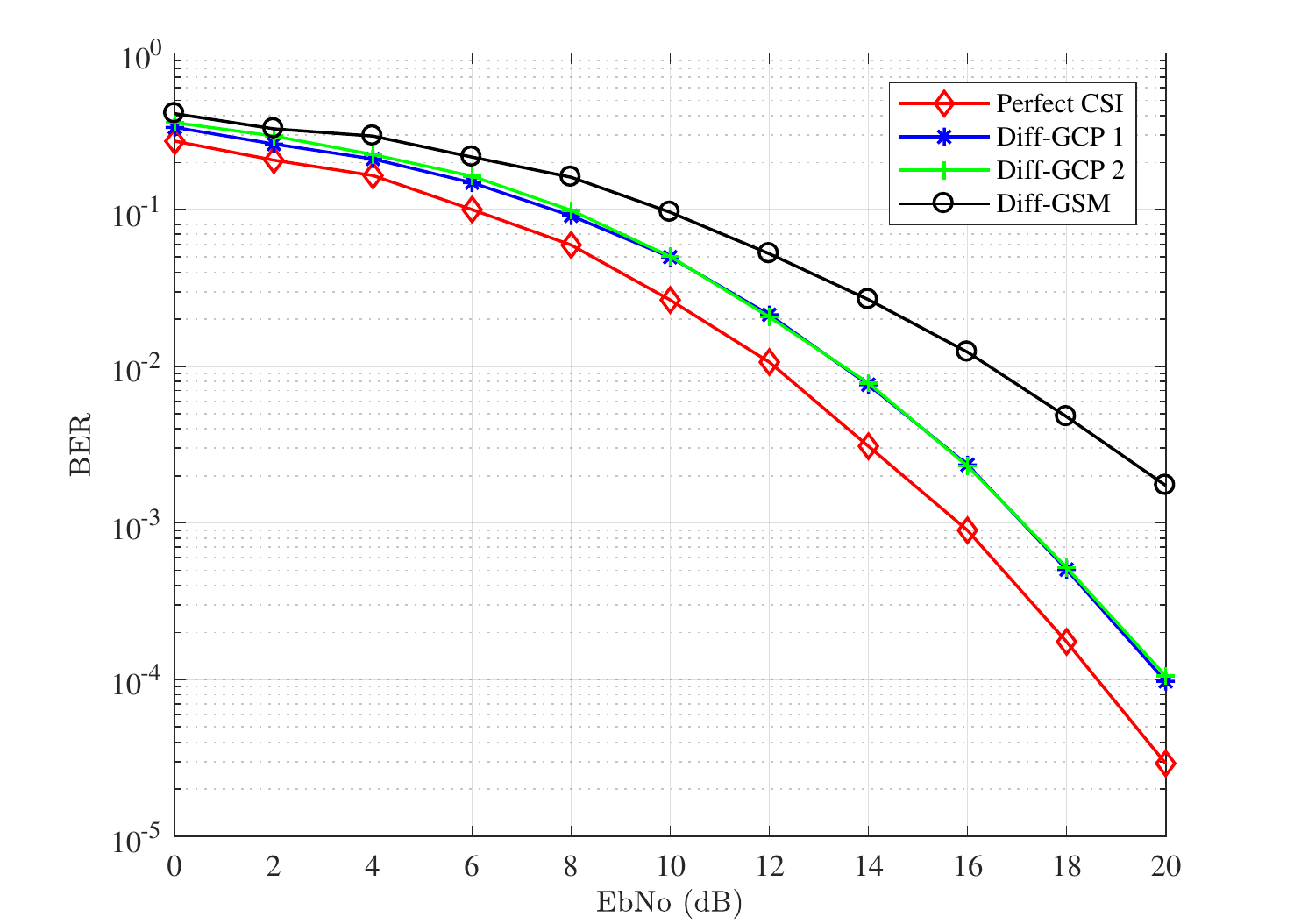}\\
  \caption{Comparison of BERs (without error-correcting codes) using different CPM training waveforms}
  \label{CPM-ChanEst-BERComp}
\end{figure}

Furthermore, under the same frequency-selective channel model, Fig. \ref{CPM-ChanEst-BERComp} compares the uncoded bit-error-rates (BERs) of GMSK systems using perfect channel state information (CSI) and estimated CSI from the proposed CPM training waveforms. Here, we follow the CPM receiver design developed in \cite{Tan2005}, where single-carrier frequency-domain equalization (SC-FDE) with minimum MSE is adopted. It is seen that the BER curves corresponding to the two Diff-GCPs are very close, and are about 1.4dB to the BER curve with perfect CSI. On the other hand, the BER curve corresponding to Diff-GSM displays 5dB distance to that with perfect CSI.

\section{Conclusions}
A systematic construction of CPM training waveform displaying autocorrelation sidelobes close to zero has been proposed. Our idea is to apply differential-encoding to a GCP and then send the encoded component sequences one after another, separated by tail bits, to the CPM modulator. Note that this work is focused on binary CPM with modulation index of $1/2$. It would be interesting to extend the presented training waveform design to generic CPM schemes (e.g., non-binary modulation orders, rational/integer modulation indices) using CPM decompositions reported in \cite{Mengali1995} and \cite{Huang2003}.
%A future work of this research is to examine the channel estimation performance of the proposed CPM training waveforms in dispersive channels.

%\section*{Acknowledgment}
%The work of Zilong Liu and Yong Liang Guan was supported by the Advanced Communications Research Program DSOCL14095, a research grant from the Future Systems and Technology Directorate, MINDEF, Singapore.

\end{document}